\documentclass[12pt]{article}%
\usepackage{fullpage}

\usepackage{orcidlink}

\usepackage[bibliography=common]{apxproof}
\usepackage{xspace}
\usepackage{amsmath,amsthm}
\usepackage{graphicx,amssymb}
\usepackage{graphics}
\usepackage{color}

\providecommand{\keywords}[1]
{
 \textbf{\textit{Keywords---}} #1
}

\newtheorem{definition}{Definition}
\newcommand{\Oh}{O}

\newtheoremrep{theorem}{Theorem}
\newtheoremrep{proposition}[theorem]{Proposition}
\newtheoremrep{lemma}[theorem]{Lemma}
\newtheoremrep{corollary}[theorem]{Corollary}

\begin{document}
\title{Polynomial Turing Compressions for Some Graph Problems Parameterized by Modular-Width}
\author{Weidong Luo
\orcidlink{0009-0003-5300-606X}\footnote{Department of Computer Science, Université de Sherbrooke, Canada} \footnote{Department of Computer Science, Humboldt-Universität zu Berlin, Germany } \footnote{\textit{Emails: weidong.luo@usherbrooke.ca or weidong.luo@yahoo.com}}}

%
\maketitle              % typeset the header of the contribution

\begin{abstract}
A polynomial Turing compression (PTC) for a parameterized problem $L$ is a polynomial time Turing machine that has access to an oracle for a problem $L'$ such that a polynomial in the input parameter bounds each query. Meanwhile, a polynomial (many-one) compression (PC) can be regarded as a restricted variant of PTC where the machine can query the oracle exactly once and must output the same answer as the oracle. Bodlaender et al. (ICALP 2008) and Fortnow and Santhanam (STOC 2008) initiated an impressive hardness theory for PC under the assumption coNP $\not\subseteq$ NP/poly. Since PTC is a generalization of PC, we define $\mathcal{C}$ as the set of all problems that have PTCs but have no PCs  under the assumption coNP $\not\subseteq$ NP/poly. Based on the hardness theory for PC, Fernau et al. (STACS 2009) found the first problem \textsc{Leaf Out-tree($k$)} in $\mathcal{C}$. However, very little is known about $\mathcal{C}$, as only a dozen problems were shown to belong to the complexity class in the last ten years. Several problems are open, for example, whether CNF-SAT($n$) and $k$-path are in $\mathcal{C}$, and novel ideas are required to better understand the fundamental differences between PTCs and PCs.

In this paper, we enrich our knowledge about $\mathcal{C}$ by showing that several problems parameterized by modular-width ($mw$) belong to $\mathcal{C}$. More specifically, exploiting the properties of the well-studied structural graph parameter $mw$, we demonstrate 17 problems parameterized by $mw$ are in $\mathcal{C}$, such as \textsc{Chromatic Number($mw$)} and \textsc{Hamiltonian Cycle($mw$)}. In addition, we develop a general recipe to prove the existence of PTCs for a large class of problems, including our 17 problems.
\end{abstract}

\section{Introduction}
Preprocessing, such as compression (kernelization) and Turing compression (kernelization), is a core research topic in parameterized complexity \cite{cyganbook,downeybook,fominbook}. Let $Q \subseteq \Sigma^* \times \mathbb{N}$ be a parameterized problem, and $f: \mathbb{N} \rightarrow \mathbb{N}$ be a computable function. 
A \textit{compression} for $Q$ is a polynomial-time algorithm that, given an instance $(x,k)$ of $Q$, returns an instance $l$ of a problem $L$ with length at most $f(k)$, such that $(x,k)\in Q$ if and only if $l\in L$. We say $Q$ admits a \textit{polynomial compression} (PC) if $f$ is a polynomial function. If $L$ equals $Q$, then the compression is called a \textit{kernelization}. 
Turing compression is a generalization of compression. 
A \textit{Turing compression} for $Q$ of size $f(k)$ is a polynomial-time algorithm $A$ with access to an oracle for a problem $L$ such that, for any input $(x,k)$, $A$ can decide whether $(x,k) \in Q$ 
sending queries of length at most $f(k)$ to the oracle.
%using queries to the oracle of length at most $f(k)$.
We say $Q$ has a \textit{polynomial Turing compression} (PTC) if $f$ is a polynomial function. If $L$ equals $Q$, then the Turing compression is called a \textit{Turing kernelization}. 
$Q$ is fixed-parameter tractable (FPT) if there is an $f(k)\cdot |x|^{\Oh(1)}$ algorithm deciding whether $(x,k) \in Q$. Note that a PTC for $Q$ is not sufficient for an FPT algorithm for $Q$ since $L$ can be undecidable.

The upper bounds and lower bounds for PCs have been studied extensively and 
%achieved uncountable 
a large number of results were achieved \cite{fominbook}. In particular, Bodlaender et al. \cite{DBLP:conf/icalp/BodlaenderDFH08,DBLP:conf/stoc/FortnowS08} initiated an impressive hardness theory to refute the existence of PCs for a large class of problems under the assumption coNP $\nsubseteq$ NP/poly. Since Turing compression 
%is generalized from compression
generalizes compression \cite{witteveen2019hierarchy}, it is possible that some natural problems without a PC admit a PTC. 
%to obtain PTCs for some problems without PCs. 
Guo \cite{bodlaender2008open} was the first to introduce the concept of Turing compression by asking whether some problems, such as the important and still open problem about $k$-path \cite{fominbook}, have PTCs but have no PCs unless coNP $\subseteq$ NP/poly. 
More than ten years ago, Fernau et al. \cite{DBLP:conf/stacs/FernauFLRSV09} found the first problem of this kind, %more specifically, they prove 
by showing that \textsc{Leaf Out-tree($k$)} has a PTC but has no PCs unless coNP $\subseteq$ NP/poly. However, by now, only about a dozen problems of this kind are known
\cite{ambalath2010kernelization,DBLP:journals/siamdm/BodlaenderJK14,DBLP:journals/jcss/DonkersJ21,Jansen17,jansenmarx,DBLP:journals/algorithmica/JansenPW19,Alexander,Thomass45}. 
In addition, despite a few results on the non-existence of PTCs~\cite{DBLP:journals/algorithmica/HermelinKSWW15,DBLP:journals/tcs/Luo22} and PTCs of restricted types \cite{burjons2021lower,DBLP:journals/corr/abs-2110-03279},
negative results on PTCs are much sparser and generally harder to obtain than positive results.
In fact, developing a framework for refuting the existence of PTCs (under widely believed assumptions) is a significant open problem in parameterized complexity, and is referred to as ``a big research challenge'' in the textbook \cite{fominbook}.  In order to tackle this ambitious challenge, more knowledge on PTCs is required.

% our knowledge about the negative results of PTC is even less than that of the above-mentioned positive results \cite{DBLP:journals/corr/abs-2110-03279, DBLP:journals/algorithmica/HermelinKSWW15, DBLP:journals/tcs/Luo22, witteveen2019hierarchy}, even though developing a framework for refuting PTC under some widely believed complexity consumption is one of the most significant open problems in parameterized complexity, which is called ``a big research challenge'' in the textbook \cite{fominbook}. 
%So getting more knowledge about PTC is always a goal in this research topic.

In this work, we focus on the PTC versus PC question for problems parameterized by modular-width ($mw$).
%Modular-width ($mw$) 
The modular-width is a well-studied structural parameter first proposed in \cite{DBLP:journals/mst/CourcelleMR00} and introduced into parameterized complexity in \cite{DBLP:conf/iwpec/GajarskyLO13}. Let $G=(V, E)$ be a graph.
A \textit{module} of $G$ is a subset of vertices $M \subseteq V$ such that, for every $v\in V \setminus M$, either $M\cap N(v) = \emptyset$ or $M\subseteq N(v)$. The empty set, $V$, and every singleton $\{v\}$ for $v\in V$ are the \textit{trivial modules}. $G$ is called a \textit{prime} graph if all modules of $G$ are trivial modules. 
The modular-width of $G$, denoted by $mw(G)$ or $mw$, is the number of vertices of the largest prime induced subgraph of $G$. In addition, $mw$ can also be defined as the number of children of the largest prime node of the \textit{modular decomposition tree}, whose definition can be found in the preliminaries.
Moreover, the dynamic programming technique can be used to design algorithms for some problems over modular decomposition trees in a bottom-up fashion. The solution of each node is obtained by combining the partial solutions of its children, where the small number of children for each node leads to an efficient algorithm. The usage of this technique can be dated back to the 1980s \cite{novick1989fast}, where some efficient parallel algorithms for problems such as \textsc{Clique}, \textsc{Max-cut}, and \textsc{Chromatic Number} are provided. In fact, from the perspective of parameterized complexity, which appears after that, the algorithms in \cite{novick1989fast} for \textsc{Clique($mw$)} and \textsc{Chromatic Number($mw$)} are FPT, and the algorithm in \cite{novick1989fast} for \textsc{Max-cut($mw$)} is XP\footnote{polynomial-time algorithm for any fixed parameter}. Recently, the technique is also used in designing FPT algorithms for graph problems in structural parameters \cite{DBLP:journals/algorithmica/BelmonteHLOO20,DBLP:journals/mst/CourcelleMR00,DBLP:conf/esa/KratschN18}.
Observe that the process of combining the partial solutions in the dynamic programming over a modular decomposition tree can be replaced by a query to an oracle with length at most a function of the largest degree of the tree. Thus, we can use Turing compression to solve a problem by simulating the dynamic programming process over the modular decomposition tree for the problem. Consequently, this technique can also help us to obtain PTCs for graph problems parameterized by $mw$, all of which coincidentally have no PCs unless coNP $\subseteq$ NP/poly.

\textbf{Our results.}
Exploiting the well-studied technique of dynamic programming algorithm over modular decomposition trees, we provide PTCs for 17 fundamental graph problems parameterized by $mw$, which have no PCs unless coNP $\subseteq$ NP/poly (some of the PC lower bounds are provided in \cite{Tablepaper}). Thus, we largely enrich the class of problems that admit PTC but do not admit PC. 
In addition, by capturing the characteristics of constructing PTCs using the technique of dynamic programming algorithm over modular decomposition tree, we develop a recipe to facilitate the development of PTCs for a large class of problems, including all the 17 problems.
In particular, our study gives rise to the following result.
\begin{theorem}
\label{theorem-ptc}
The following problems parameterized by $mw$ have PTCs but have no PCs unless coNP $\subseteq$ NP/poly:
\textsc{Independent Set, Clique}, \textsc{Vertex Cover},  
\textsc{Chromatic Number, Dominating Set}, 	 
\textsc{Hamiltonian Cycle, Hamiltonian Path}, 
\textsc{Feedback Vertex Set}, \textsc{Odd cycle Transversal}, \textsc{Connected Vertex Cover}, \textsc{Induced Matching}, \textsc{Nonblocker}, \textsc{Maximum Induced Forest}, \textsc{Partitioning Into Paths}, \textsc{Longest Induced Path}, \textsc{Independent Triangle Packing}, and \textsc{Independent Cycle Packing},
where the results of the PC lower bounds for the first 11 problems are demonstrated in \cite{Tablepaper}.
\end{theorem}

%\noindent \textbf{Orgnization.} Section 2 is the preliminaries. Section 3 provides a recipe for obtaining PTC in the parameter $mw$. Section 4 proves the PTCs for all the problems in this paper using the recipe. Section 5 proposes some open questions as conclusions. In addition, the Appendix  demonstrates the nonexistence of PCs for the last six problems of Theorem \ref{theorem-ptc}.

\section{Preliminaries}
We denote $\Sigma = \{0,1\}$ and $[n] = \{1,\ldots,n\}$. Let $\overline G$ denote the complement of a graph $G$. Unless otherwise specified, $V(G)$ and $E(G)$ indicate the vertex and edge sets of $G$, respectively. For $v \in V(G)$, $N(v)$ consists of all neighbors of $v$. We denote $N[v] = N(v) \cup \{v\}$. 
%Let $M\subseteq V(G)$. $G[M]$ denotes the subgraph induced in $G$ by $M$. 
For $M\subseteq V(G)$, $G[M]$ denotes the subgraph induced in $G$ by $M$, and $N(M)$ consists of all vertices that are not in $M$ but are adjacent to some vertex of $M$. We denote $N[M] = N(M) \cup M$. The cardinality of a set $S$ is denoted by $|S|$. Symbols $\mathcal{G}$ and $\mathbb{N}$ denote the sets of undirected graphs and the natural numbers, respectively. 
For two disjoint vertex sets $M$ and $M'$ of a graph, the edges between $M$ and $M'$ 
refer to
%indicate 
all edges $uv$ such that $u\in M$ and $v\in M'$. In addition, we say $M$ and $M'$ are adjacent if all possible edges between $M$ and $M'$ exist, and $M$ and $M'$ are non-adjacent if there are no edges between $M$ and $M'$.
$K_n$ and $C_n$ denote a complete graph and a cycle with $n$ vertices, respectively. For $X\subseteq V(G)$, we write $G-X$ for the subgraph induced in $G$ by $V(G)\setminus X$. In addition, we also use $G - G'$ to represent $G - V(G')$ for a subgraph $G'$ of $G$. The intersection and union of two graphs $G = (V,E)$ and $G'=(V',E')$ are denoted as $G\cap G' = (V\cap V', E\cap E')$ and $G\cup G' = (V\cup V', E\cup E')$, respectively. The $\Oh^*$-notation suppresses factors that are polynomial in the input size.

Let $G=(V, E)$ be a graph. Recall that a module of $G$ is a subset of vertices $M \subseteq V$ such that, for every $v\in V \setminus M$, either $M\cap N(v) = \emptyset$ or $M\subseteq N(v)$. A module $M$ is a \textit{strong module} if, for any module $M'$, only one of the following holds: (1) $M\subseteq M'$ (2) $M'\subseteq M$ (3) $M\cap M' = \emptyset$. A module $M$ is \textit{maximal} if $M\subsetneq V$ and no module $M'$ satisfies $M \subsetneq M' \subsetneq V$. Assume $P \subseteq 2^V$ is a vertex partition of $V$. $P$ is a \textit{(maximal) modular partition} if all $M\in P$ are (maximal strong) modules of $G$. 
For a modular partition $P$, the \textit{quotient graph} $G_{/P} =(V_P, E_P)$ is defined as follows. 
The set of vertices $V_P$ contains one vertex $v_M$ for each module $M \in P$, so that $V_P = \{v_M : M \in P\}$. An edge $v_{M}v_{M'}$ is contained in $E_P$ if and only if $M,M'\in P$ are adjacent. 
All strong modules $M$ of $G$ can be represented by an inclusion tree $MD(G)$, where each $M$ corresponds to a vertex $v_M$ of $MD(G)$, and, for any two strong modules $M, M'$ of $G$, $v_{M'}$ is a descendant of $v_M$ in $MD(G)$ if and only if $M' \subsetneq M$. This unique tree $MD(G)$ is called the \textit{modular decomposition tree} of $G$. 
The internal vertices are divided into three types: a vertex $v_M$ is \textit{parallel} if $G[M]$ is disconnected, \textit{series} if $\overline{G[M]}$ is disconnected, \textit{prime} if both $G[M]$ and $\overline{G[M]}$ are connected.
%All the internal vertices $v_M$ of $MD(G)$ are divided into three types: parallel, series, and prime. $v_M$ is \textit{parallel} if $G[M]$ is disconnected. $v_M$ is \textit{series} if $\overline{G[M]}$ is disconnected. $v_M$ is \textit{prime} if both $G[M]$ and $\overline{G[M]}$ are connected. 
The \emph{modular-width} of $G$ can also be defined as the minimum number $k$ such that the number of children of any prime vertex in $MD(G)$ is at most $k$. 
In addition, for a module $M$ of $G$, $G[M]$ is called a \textit{factor}. The modular decomposition tree of $G$ can be obtained in time $O(m+n)$ \cite{Tedder_2008}. Refer to \cite{HabibP10} for more information about modular decomposition trees. 
The null graph and the empty module are disregarded in the proofs of this paper (the results are trivial for these cases).

Next, we give the definitions of the problems of Theorem \ref{theorem-ptc}, all of which are NP-hard. Given a graph $G$, \textsc{Hamiltonian Cycle} (\textsc{Hamiltonian Path}) asks whether $G$ has a cycle (path) that visits each vertex of $G$ exactly once. Let $(G,k)$ be the input of the following problems, where $G=(V,E)$ is a graph and $k$ is an integer. \textsc{Chromatic Number} asks whether $V$ can be colored by at most $k$ colors such that no two adjacent vertices share the same color. \textsc{Clique} asks whether $V$ has a subset of size at least $k$ such that any two vertices in it are adjacent. \textsc{Vertex Cover} asks whether $V$ has a subset, called vertex cover, of size at most $k$ such that every edge of $G$ has at least one endpoint in it. \textsc{Connected Vertex Cover} asks whether $G$ has a vertex cover $X$ such that $|X| \leq k$ and $G[X]$ is connected. \textsc{Dominating Set} asks whether $V$ has a subset of size at most $k$ such that every vertex not in it is adjacent to at least one vertex of it. \textsc{Feedback Vertex Set} asks whether $V$ has a subset $X$ of size at most $k$ such that $G - X$ is a forest. \textsc{Independent Cycle (Triangle) Packing} asks whether $G$ contains an induced subgraph consisting of at least $k$ pairwise vertex-disjoint cycles (triangles). \textsc{Independent Set} asks whether $V$ has a subset of size at least $k$ such that any two vertices in it are not adjacent. \textsc{Induced Matching} asks whether $V$ has a subset $X$ of size at least $2k$ such that $G[X]$ is a matching with at least $k$ edges. \textsc{Longest Induced Path} asks whether $G$ contains the path on $k$ vertices as an induced subgraph. \textsc{Max Leaf Spanning Tree} asks whether $G$ has a spanning tree with at least $k$ leaves. \textsc{Nonblocker} asks whether $V$ has a subset of size at least $k$ such that every vertex in it is adjacent to a vertex outside of it. \textsc{Odd Cycle Transversal} asks whether $V$ has a subset $X$ of size at most $k$ such that $G - X$ is a bipartite graph. \textsc{Partitioning Into Paths} asks whether $G$ contains $k$ vertex disjoint paths whose union includes every vertex of $G$. The function (optimization) versions of all the above-mentioned problems are defined in natural ways, for example, for the function (optimization) version of \textsc{Clique}, the input is $G$ and the output is the number of vertices of the largest clique of $G$.

\section{Recipe for polynomial Turing compression in parameter modular-width}\label{recipesecton}

Suppose we are given graphs $G=(V,E)$ and $H=(V_H,E_H)$, as well as a module $M$ of $G$. Assume $G_S$ is a supergraph of $G$, which is obtained as follows: (1) add $G$ and $H$ into $G_S$, (2) add $uv$ to $G_S$ for all $v\in N(M), u\in V_H$. Let $G'$ be the subgraph induced  in $G_S$ by $(V\setminus M)  \cup V_H$. We say $G'$ is obtained from $G$ by replacing $G[M]$ with $H$, and the process of obtaining $G'$ from $G$ is called a \textit{modular replacement}. Clearly, $V_H$ is a module of $G'$. 
Recall that symbols $\mathcal{G}$ and $\mathbb{N}$ denote the sets of undirected graphs and the natural numbers, respectively.

\begin{lemma}
\label{well defined lemma}
Let each $F_i$ be a function from $\mathcal{G}$ to $\mathbb{N}$ for $i \in [r]$. For any graphs $G$ and $H$, as well as any module $M$ of $G$, suppose $F_i(G[M])=F_i(H)$ for all $i$ implies $F_i(G)=F_i(G')$ for all $i$, where $G'$ is obtained from $G$ by replacing $G[M]$ with $H$. 
Then, for any graph $G$ and any modular partition $P$ of $V(G)$, the quotient graph $G_{/P}$ together with $F_1(G[M]),\ldots,F_r(G[M])$ for all modules $M$ in $P$ completely determine $F_1(G),\ldots,F_r(G)$.
\end{lemma}

\begin{proof}
Let tuple $T(G) = (F_1(G),\ldots,F_r(G))$. For a graph $G$ and a modular partition $P$ of $V(G)$, we say $X$ is a \emph{values-attached quotient graph} generated from $P$ if $X$ is the quotient graph $G_{/P}=(V_P,E_P)$ with each vertex $v_M \in V_P$ attached the tuple $T(G[M])$. 
Suppose $\mathcal{X}$ consists of all possible $X$ generated from any $P$ of $V(G)$, where $G\in \mathcal{G}$. Assume the binary relation $R$ over $\mathcal{X}$ and $\mathcal{G}$ consists of all pairs $(X, G)$ such that $X$ is generated from some modular partition of $V(G)$. For each $i \in [r]$, let binary relation $f_i$ be the composition relation of $R$ and $F_i$ over $\mathcal{X}$ and $\mathbb{N}$, which means that $f_i = R;F_i = \{(X,n) \mid \text{there exists } G \in \mathcal{G} \text{ such that } (X,G) \in R \text{ and }  (G,n) \in F_i \}$.

Consider every $f_i$. According to the definition of $R$, for any $X \in \mathcal{X}$, there is at least a $G\in \mathcal{G}$ such that $(X, G) \in R$. Moreover, as $F_i$ is a function, for any $G\in \mathcal{G}$, there is an $n\in \mathbb{N}$ such that $(G,n) \in F_i$. Therefore, for any $X \in \mathcal{X}$, there is at least an $n\in \mathbb{N}$ such that $(X,n)\in f_i$, so $f_i$ is left-total.
For an $X\in \mathcal{X}$, assume $\mathcal{G'}$ consists of all $G$ such that $(X,G) \in R$. We claim that $F_i(G_1) = F_i(G_2)$ for any $G_1, G_2 \in \mathcal{G'}$. Since $(X,G_1), (X,G_2) \in R$, there exist modular partitions $P_1$ of $G_1$ and $P_2$ of $G_2$ such that $X$ is not only generated from $P_1$ but also generated from $P_2$.
Thus, the quotient graphs $G_{1/P_1}$ and $G_{2/P_2}$ are isomorphic, and there exists an edge-preserving bijection $g$ from $V(G_{1/P_1})$ to $V(G_{2/P_2})$ such that $T(G[M_1])=T(G[M_2])$ for every $v_{M_1}$ of $V(G_{1/P_1})$ and $v_{M_2} = g(v_{M_1})$ of $V(G_{2/P_2})$. Here, we say $G[M_2]$ of $G_2$ corresponds to $G[M_1]$ of $G_1$ if $v_{M_2} = g(v_{M_1})$. 
Now, consider every module $M_1$ of $P_1$. We replace each $G[M_1]$ of $G_1$ with its corresponding factor $G[M_2]$ of $G_2$ one by one. According to the prerequisite of this lemma,\footnote{The prerequisite of this lemma is the second sentence of Lemma \ref{well defined lemma}.} since $T(G[M_1]) = T(G[M_2])$,  the value of $F_i$ for the new obtained graph after every modular replacement does not change. After the final modular replacement, $G_2$ is obtained, so we have $F_i(G_1) = F_i(G_2)$. Hence, $F_i(G)$ is a fixed number for any $G\in \mathcal{G'}$. 
This means that $f_i$ is right-unique 
(note that a relation is called right-unique if each element on the right side of the relation is mapped to a unique element on the left side). 
Consequently, $f_i$ is a function from $\mathcal{X}$ to $\mathbb{N}$, moreover, $F_i(G)$ equals $f_i(X)$, where $(X,G) \in R$. 
\end{proof}

%\ml{[NOTE ML: I suggest adding a $\qed$ at the end of each proof.]}

Note that, for each $i$, we say an algorithm solves $F_i(G)$ if it outputs $F_i(G)$ with the input $G$. The decision version of $F_i(G)$ is as follows: given a graph $G$ and an integer $k$, decide whether $F_i(G)\leq k$ (or $F_i(G)\geq k$). 
%Recall that $mw$ denotes the modular-width of $G$.

\begin{lemma}
\label{core PTK lemma}	
Let each $F_i$ be a function from $\mathcal{G}$ to $\mathbb{N}$ for $i \in [r]$, where $r$ is a constant. Assume the following statements hold.
\begin{enumerate}
\item 
For any graphs $G$ and $H$, as well as any module $M$ of $G$, $F_i(H)=F_i(G[M])$ for all $i$ implies $F_i(G) \leq F_i(G')$ (or $F_i(G) \geq F_i(G')$) for all $i$,\footnote{Here, we only require every $F_i$ is monotone under module-substitution.} where $G'$ is obtained from $G$ by replacing $G[M]$ with $H$.
\item 
For each $i$, there is a $2^{mw^{\Oh(1)}}|G|^{\Oh(1)}$ time algorithm to solve $F_i(G)$.
\item 
$F_i(G) \leq |G|^{\Oh(1)}$ for each $i$.
\end{enumerate}
Then each $F_i(G)$ can be solved by a polynomial-time algorithm together with an oracle for a problem $Q$ that can decide in one step whether a string of length $mw^{\Oh(1)}$ is in $Q$.
Moreover, the decision version of each $F_i(G)$ has a PTC parameterized by $mw$.
\end{lemma}

\begin{proof}
For all $i$, assume $F_i(G) \leq |G|^c$ and $F_i(G)$ can be solved in $2^{mw^c}|G|^{c}$ for some constant $c$. Suppose w.l.o.g. that $mw \geq \log^{\frac{1}{c}} |G|$ henceforth (otherwise, each $F_i(G)$ can be solved in polynomial time based on statement 2 and the consequence of this lemma holds). 
Consider statement 1. Since $V(H)$ is a module of $G'$, we can obtain $F_i(G) = F_i(G')$ by exchanging the position of $G$ and $G'$ as follows. For the graphs $G'$ and $G[M]$, as well as the module $V(H)$ of $G'$, that $F_i(G[M])=F_i(H)$ for all $i$ also implies that $F_i(G) \geq F_i(G')$ (or $F_i(G) \leq F_i(G')$) for all $i$, where $G$ is obtained from $G'$ by replacing $H$ with $G[M]$. 
Thus, functions $F_1,\ldots,F_r$ fulfill the conclusion of Lemma \ref{well defined lemma}. Let the definitions of values-attached quotient graph $X$, set $\mathcal{X}$, binary relation $R$ and functions $f_1, \ldots, f_r$ be the same as that in the proof of Lemma \ref{well defined lemma}. Assume $Q$ consists of all the strings $(X,f_1(X),\ldots,f_r(X))$ for $X\in \mathcal{X}$.

Recall that every vertex $v_M$ of the modular decomposition tree $MD(G)$ of a graph $G$ corresponds to the strong module $M$ of $G$, where $MD(G)$ can be constructed in linear time. Here, we call $G[M]$ the corresponding graph of $v_M$. 
Roughly speaking, to obtain $F_1(G),\ldots, F_r(G)$, we compute in a bottom-up fashion the values of all $F_i$ for the graphs that correspond to the vertices of $MD(G)$.
First, the corresponding graph of each leaf of $MD(G)$ is the singleton graph $K_1$, where $F_i(K_1)$ can be solved in $\Oh(1)$ time according to statement 2.
Secondly, consider an internal vertex $v_M$ of $MD(G)$. Let $P$ be the maximal modular partition of $M$. Then every child $v_{M'}$ of $v_M$ corresponds to the module $M'$ of $P$ and the quotient graph of $v_M$ is $G[M]_{/P}$.
Suppose $v_M$ is prime. 
Assume $X \in \mathcal{X}$ is the values-attached quotient graph generated from $P$. Since $F_1(G[M']),\ldots,F_r(G[M'])$ for all $M'$ of $P$ and $G[M]_{/P}$ are given, $X$ can be obtained immediately. Let $k_1,\ldots,k_r$ be non-negative integers at most $|G|^c$. Exhaustively generate string $(X,k_1,\ldots,k_r)$ and query the oracle whether it is in $Q$. Based on Lemma \ref{well defined lemma}, we have $f_i(X)=F_i(G[M]) \leq |G|^c$ for all $i$. Hence, we can obtain $F_1(G[M]),\ldots,F_r(G[M])$ by finding the string $(X,k_1,\ldots,k_r)$ $=(X,f_1(X),\ldots,f_r(X))$ after querying the oracle at most $(|G|^c+1)^r=|G|^{\Oh(1)}$ times.
Moreover, the length of any $(X,k_1,\ldots,k_r)$ is at most  $\Oh(mw^2\log|G|) = mw^{\Oh(1)}$.
Assume $v_M$ is parallel. Suppose $P$ contains $t$ modules, where $2\leq t\leq |V(G)|$. Then, the quotient graph $G[M]_{/P}$ is $\overline K_t$. 
Consider any two modules $M'_1$ and $M'_2$ of $P$. Let $M'_{12}$ be $M'_1 \cup M'_2$ and $G[M'_{12}]$ be the subgraph induced by $M'_{12}$ in $G[M]$. Then, $P'=\{M_1',M_2'\}$ is a modular partition of $M'_{12}$. Since the values-attached quotient graph generated from $P'$ are given, we can obtain $F_1(G[M'_{12}]),\ldots,F_r(G[M'_{12}])$ using the same method as that of the prime vertex.
Now, consider the new modular partition $P=\{M'_{12}\} \cup (P \setminus \{M_1',M_2'\})$ of $M$. It contains only $t-1$ modules and the new quotient graph $G[M]_{/P}$ is $\overline K_{t-1}$.  Moreover, $F_1(G[M']),\ldots,F_r(G[M'])$ are known for every module $M'$ in $P$. 
Clearly, this process decreases the vertex number of the quotient graph by one. We can repeat $t-1$ times the same process on every newly generated quotient graph. Finally, $F_1(G[M]),\ldots, F_r(G[M])$ can be obtained. 
Assume $v_M$ is a series vertex. $F_1(G[M]),\ldots, F_r(G[M])$ can be obtained using the same strategy as that of the parallel vertex.
Therefore, for any internal vertex $v_M$ of $MD(G)$, we obtain $F_1(G[M]),\ldots, F_r(G[M])$ in $|G|^{\Oh(1)}$ time with each query length at most $mw^{\Oh(1)}$. 
In addition, the vertex number of $MD(G)$ is $O(|G|)$. As a result, $F_1(G),\ldots, F_r(G)$ can be solved by a polynomial-time algorithm with an oracle for $Q$ that can decide whether a string of length $mw^{\Oh(1)}$ is in $Q$. Furthermore, according to the definition of PTC, the decision version of each $F_i(G)$ has a PTC parameterized by $mw$.
\end{proof}

\section{Polynomial Turing compressions for problems}
Obviously, the function versions of all problems in Theorem \ref{theorem-ptc} fulfill statement 3 of Lemma \ref{core PTK lemma}. 
\textsc{Clique}, \textsc{Feedback Vertex Set}, \textsc{Longest Induced Path}, \textsc{Induced Matching}, \textsc{Independent Triangle Packing}, \textsc{Independent Cycle Packing} can be solved in $\Oh^*(1.74^{mw})$ time \cite{Fomin_2017}. \textsc{Chromatic Number} can be solved in $\Oh^*(2^{mw})$ time \cite{DBLP:conf/iwpec/GajarskyLO13}. \textsc{Hamiltonian Cycle} and \textsc{Partitioning Into Paths} can be solved in $2^{\Oh(mw^2\log mw)}n^{\Oh(1)}$ time \cite{DBLP:conf/iwpec/GajarskyLO13}. In addition,
clique-width ($cw$) \cite{DBLP:journals/dam/CourcelleO00} is a generalized parameter of $mw$ such that $cw \leq mw+2$ in a graph (the $cw$ and $mw$ of a cograph are two and zero, respectively). \textsc{Connected Vertex Cover} \cite{DBLP:journals/tcs/BergougnouxK19}, \textsc{Dominating Set} \cite{DBLP:conf/mfcs/BodlaenderLRV10}, and \textsc{Odd Cycle Transversal} \cite{hegerfeld2022towards,jacob2021close} can be solved in $2^{\Oh(cw)}n^{\Oh(1)}$ time, thus also in $2^{\Oh(mw)}n^{\Oh(1)}$ time. Therefore, the function versions of all the above-mentioned problems fulfill statement 2 of Lemma \ref{core PTK lemma}.

Clearly, a PTC of each problem in Theorem \ref{theorem-ptc} can be obtained if we can obtain a PTC of the problem with connected input graphs. So we assume w.l.o.g. the input graph of every problem is connected. 
In this section, unless otherwise specified, assume $G=(V, E)$ and $H=(V_H,E_H)$ are connected graphs, $M\neq \emptyset$ is a module of $G$, $M' = N(M)$ in $G$, and graph $G'=(V',E')$ is obtained from $G$ by replacing $G[M]$ with $H$. 
Now, we only need to prove the function versions of the above-mentioned problems, all of which will be discussed in this section, fulfill statement 1 of Lemma \ref{core PTK lemma} to provide PTCs for the problems parameterized by $mw$. 
More specifically, we will prove that $F_i(H)=F_i(G[M])$ for all $i$ implies that $F_i(G) \leq F_i(G')$ (or $F_i(G) \geq F_i(G')$) for all $i$, where functions $F_i$ are the function versions of the problems discussed in some lemma of this section. Obviously, the statement is true for any function $F_i$ if $M = V$. So assume $M \neq V$ henceforth. In addition, $M'\neq \emptyset$ since $G$ is connected and $M\not \in \{ \emptyset, V\}$.
In this section, assume function $F_{v} (I)$ denotes the vertex number of $I$ for any $I \in \mathcal{G}$.
Let min-DS, min-CVC, min-VC, min-FVS, min-OCT, max-IM, max-ITP, and max-ICP be the abbreviations of the minimum dominating set, minimum connected vertex cover, minimum vertex cover, minimum feedback vertex set, minimum odd cycle transversal, maximum induced matching, maximum independent triangle packing, and maximum independent cycle packing, respectively.

%\begin{lemma}\label{clique-ptc}
%\textsc{Clique($mw$)} has a PTC.
%\end{lemma}
%\begin{proof}
%Let function $F (I)$ denote the clique number for any $I \in \mathcal{G}$. Assume $F (G[M])=F (H)$. Let $C$ and $C_H$ be maximum cliques of $G$ and $H$, respectively. Assume $C\cap M = \emptyset$. Clearly, $F (G') \geq F (G)$. Assume $C\cap M \neq \emptyset$. Then, $C\setminus M$ is adjacent to $M$ in $G$, thus also adjacent to $V_H$ in $G'$. Hence, $(C\setminus M)\cap C_H$ with size $F(G)$ is a clique of $G'$, and $F (G') \geq F (G)$. 
%\end{proof}

\begin{lemma}
\label{chromatic}
\textsc{Chromatic Number($mw$)} has a PTC.
\end{lemma}

\begin{proof}
Let function $F (I)$ denote the chromatic number  for any $I \in \mathcal{G}$.
Suppose $F (G[M])=F (H)$. Then, there is a coloring $c : V \rightarrow C$ for $G$, where $C=[F(G)]$. Let $C_{M} = \{c(v) \mid v\in M\}$ and $C_{M'} = \{c(v) \mid v\in M'\}$.  Since $M' = N(M)$, $C_{M} \cap C_{M'} = \emptyset$.
Consider $G'$. Since $F (H)=F (G[M]) \leq |C_{M}|$, there is a coloring $c_H : V_H \rightarrow C_{M}$ for $H$. Suppose $c' : V' \rightarrow C$ is a function such that $c'(v)=c(v)$ for all $v\in V'\setminus V_H$ and $c'(v)=c_H(v)$ for all $v\in V_H$. Since $N(V_H)=M'$ and $C_{M} \cap C_{M'} = \emptyset$, $c'$ is a coloring for $G'$. Thus, $F (G')\leq |C| = F (G)$.
\end{proof}

\begin{lemma}
\label{dslemma2}
Let $D$ be a min-DS of $G$. Then $M\cap D$ is either $\emptyset$, $\{v\}$, or a min-DS of $G[M]$.
\end{lemma}
\begin{proof}
Assume, for contradiction, $|D \cap M| \geq 2$ and $D \cap M$ is not a min-DS of $G[M]$. If $D \cap M$ is a dominating set of $G[M]$ that is not minimum, then let $X$ be a smaller one, and $(D \setminus M) \cup X$ is a smaller dominating set of G, a contradiction. Hence, $D \cap M$ is not a dominating set of $G[M]$, so there must be an $x \in D\cap M'$. Then $M \subseteq N(x)$ and every vertex in $M$ has the same neighborhood outside of $M$, so $D$ is still a dominating set of $G$ by removing from $D$ all but one vertex of $M \cap D$, contradicting the minimality of $D$.
\end{proof}

\begin{lemma}
\label{dominating set 222}
\textsc{Dominating Set($mw$)} has a PTC.
\end{lemma}
\begin{proof}
Let function $F(I)$ denote the size of the min-DS for any $I \in \mathcal{G}$. Suppose $F (G[M])=F (H)$ and $D$ is a min-DS of $G$. 
Assume $D\cap M = \emptyset$. Then there exists an $x\in D\cap M'$ such that $V_H \subseteq N(x)$ in $G'$.  Thus, $D$ is a dominating set of $G'$ and $F(G')\leq F(G)$.	
Assume $D\cap M = \{u\}$. Suppose $F (G[M])=F (H) \geq 2$. Then $\{u\}$ is not a dominating set of $G[M]$, so there is a $v\in D\cap M'$ such that $V_H \subseteq N(v)$ in $G'$. Clearly, $\{w\}\cup (D\setminus \{u\})$ is a dominating set of $G'$ for any $w\in V_H$, so $F(G')\leq F(G)$. Suppose $F (G[M])=F (H) =1$. We may assume $\{v\}$ is a dominating set of $H$. Clearly, $\{v\}\cup (D\setminus \{u\})$ is a dominating set of $G'$, so $F(G')\leq F(G)$.
Now, according to Lemma \ref{dslemma2}, we only need to consider that $D\cap M$ is a min-DS of $G[M]$. Suppose $D_H$ is a min-DS of $H$. Clearly, $D_H\cup (D\setminus M)$ is a dominating set of $G'$, so $F(G')\leq F(G)$.
\end{proof}

\begin{lemma}
\label{vclemma}
Assume $I_{G[M]}$ and $I_H$ with the same size are independent sets of $G[M]$ and $H$, respectively. Suppose $S\subseteq V \setminus M$. Then, the subgraph in $G$ induced by $S \cup I_{G[M]}$ and the subgraph in $G'$ induced by $S \cup I_H$ are isomorphic.
\end{lemma}
\begin{proof}
It is trivial if $S$ or $I_{G[M]}$ is empty. 
Assume $I_{G[M]}$ and $S$ are not empty. Since $|I_H| = |I_{G[M]}|$, we may assume $S=\{v_1,\ldots, v_r\}$, $I_{G[M]}=\{u_1,\ldots, u_s\}$, and $I_H = \{w_1,\ldots, w_s\}$. Suppose $f$ is a bijection from $S \cup I_{G[M]}$ to $S \cup I_H$ such that $f=\{[v_1,v_1],\ldots,[v_r,v_r],[u_1,w_1],\ldots,[u_s,w_s]\}$. Clearly, $f$ is an edge-preserving bijection.
\end{proof}

\begin{lemma}
\label{noedge}
$G$ has an edge if and only if $G[V\setminus S]$ has an edge, where $S$ is a min-FVS or a min-OCT of $G$.
\end{lemma}
\begin{proof}
Let $S$ be a min-FVS of $G$. For the forward direction, suppose $G$ has an edge. $G[V\setminus S]$ has an edge if $S=\emptyset$. 
Assume $S\neq \emptyset$. After deleting any $|S|-1$ vertices of $S$ from $G$, there exists a cycle in $G$, otherwise, $G$ has a feedback vertex set of size $|S|-1$. Hence, $G[V\setminus S]$ has at least one edge since the edges of a cycle cannot be entirely removed by deleting one vertex. The reverse direction is trivial.
The proof goes the same way if $S$ is a min-OCT of $G$
\end{proof}

\begin{lemma}
\label{octlemma2}
Assume $C, F, O, R$ are a min-VC, a min-FVS, a min-OCT, and a min-CVC of $G$, respectively. Let $v$ be a vertex of $M$.
The following statements hold.
(1) $M \cap C$ is either $M$ or a min-VC of $G[M]$.
(2) $M \cap F$ is either $M$, $M \setminus \{v\}$,  a min-VC of $G[M]$, or a min-FVS of $G[M]$.
(3) $M \cap O$ is either $M$, a min-VC of $G[M]$, or a min-OCT of $G[M]$.
(4) $M \cap R$ is either $M$, a min-VC of $G[M]$, or $\{v\}$.
%(4) If $G[M]$ has an edge, then $M \cap C$ is either $M$ or a min-VC of $G[M]$, otherwise, $M\cap C$ is either $M$, a min-VC of $G[M]$, or $\{v\}$ for a $v \in M$.
\end{lemma}
\begin{proof}
Recall that $M' \neq \emptyset$. (1) If $M \cap C\neq M$, then $M'\subseteq C$ and $M \cap C$ is a min-VC of $G[M]$. 
(2) Assume $G[M\setminus F]$ has an edge. Then $M'\subseteq F$ and $M\cap F$ is a min-FVS of $G[M]$. Assume $G[M\setminus F]$ has no edges but has at least two vertices. Then $M'\setminus F$ contains at most one vertex, otherwise, there exists a $C_4$ in $G[V\setminus F]$. Hence, $F\cap M$ is a min-VC of $G[M]$. Assume $M\setminus F$ equals $\{v\}$ or $\emptyset$. Then $M \cap F$ is  $M \setminus \{v\}$ or $M$.
(3) %Suppose $M'= \emptyset$. Then $M\cap O$ is a min-OCT of $G[M]$. Suppose $M' \neq \emptyset$. 
Clearly, $M\cap O = M$ if $M\setminus O = \emptyset$. Assume $G[M\setminus O]$ contains an edge. Then $M'\subseteq O$, so $M\cap O$ is a min-OCT of $G[M]$. Assume $G[M\setminus O]$ contains a vertex but no edges. Then $G[M'\setminus O]$ contains no edges, so $M \cap O$ is a min-VC of $G[M]$.  
(4) Clearly, $M \subseteq R$ if $M' \not\subseteq R$. Assume $M'\subseteq R$ henceforth. Then, $M\cap R$ is a min-VC of $G[M]$ if $G[R\setminus M]$ is connected or $G[M]$ contains an edge, otherwise, $M \cap R$ is a vertex of $M$ to ensure the connectivity of $G[R]$.
\end{proof}

%\begin{lemma}
%\label{cvclemma1}
%Suppose $C$ is a min-CVC of $G$. If $G[M]$ has an edge, then $M \cap C$ is either $M$ or a min-VC of $G[M]$, otherwise, $M\cap C$ is either $M$, a min-VC of $G[M]$, or $\{v\}$ for a $v \in M$.
%\end{lemma}
%\begin{proof}
%Clearly, $M \subseteq C$ if $M' \not\subseteq C$. Assume $M'\subseteq C$ henceforth. Then, $M\cap C$ is a min-VC of $G[M]$ if $G[C\setminus M]$ is connected or $G[M]$ has an edge, otherwise, $M \cap C$ is a vertex of $M$ to ensure the connectivity of $G[C]$.
%\end{proof}

%A trivial connected component of a graph is a singleton. We assume w.l.o.g. that the input graph of \textsc{Connected Vertex Cover($mw$)} is connected. Otherwise, the CVC of a graph does not exist if it includes two non-trivial connected components, and the CVC of a graph is the CVC of its largest connected component if it includes at most one non-trivial connected component.

\begin{lemma}
\label{OCT}
\textsc{Vertex Cover($mw$)}, \textsc{Connected Vertex Cover($mw$)}, \textsc{Feedback Vertex Set($mw$)}, and \textsc{Odd Cycle Transversal($mw$)} have PTCs.
\end{lemma}
%\begin{proofsketch}
%Suppose functions $F_{oct}(I)$, $F_{fvs}(I)$, $F_{cvc}(I)$, and $F_{vc}(I)$ represent the sizes of min-OCT, min-FVS, min-CVC, and min-VC of any $I \in \mathcal{G}$, respectively. 
%Suppose $F_{v}(G[M])=F_{v}(H)$, $F_{vc}(G[M])=F_{vc}(H)$, $F_{cvc}(G[M])=F_{cvc}(H)$, $F_{fvs}(G[M])=F_{fvs}(H)$, and $F_{oct}(G[M])=F_{oct}(H)$.
%Let $C$, $R$, $F$, and $O$ represent a min-VC, a min-CVC, a min-FVS, and a min-OCT of $G$, respectively.
%Let $C_H$, $F_H$, and $O_H$ represent a min-VC, a min-FVS, and a min-OCT of $H$, respectively. Obviously, $F_{v}(G') \leq F_{v}(G)$. Moreover, we can prove $F_{vc}(G') \leq F_{vc}(G)$, $F_{cvc}(G') \leq F_{cvc}(G)$,  $F_{fvs}(G') \leq F_{fvs}(G)$, and $F_{oct}(G') \leq F_{oct}(G)$ using Lemma \ref{vclemma}, \ref{noedge}, and \ref{octlemma2}.
%\end{proofsketch}

\begin{proof}
Suppose functions $F_{oct}(I)$, $F_{fvs}(I)$, $F_{cvc}(I)$, and $F_{vc}(I)$ represent the sizes of min-OCT, min-FVS, min-CVC, and min-VC of any $I \in \mathcal{G}$, respectively. 
Suppose $F_{v}(G[M])=F_{v}(H)$, $F_{vc}(G[M])=F_{vc}(H)$, $F_{cvc}(G[M])=F_{cvc}(H)$, $F_{fvs}(G[M])=F_{fvs}(H)$, and $F_{oct}(G[M])=F_{oct}(H)$.
Let $C$, $R$, $F$, and $O$ represent a min-VC, a min-CVC, a min-FVS, and a min-OCT of $G$, respectively.
Let $C_H$, $F_H$, and $O_H$ represent a min-VC, a min-FVS, and a min-OCT of $H$, respectively. Obviously, $F_{v}(G') \leq F_{v}(G)$. 

We claim $F_{vc}(G') \leq F_{vc}(G)$.
Based on Lemma \ref{octlemma2}, $C\cap M$ is either $M$ or a min-VC of $G[M]$. 
Let $S$ denote $V \setminus (C\cup M)$.
Assume $C\cap M = M$. Clearly, $(C\setminus M) \cup V_H$ is a vertex cover (VC) of $G'$. Assume $C\cap M$ is a min-VC of $G[M]$. Then, $M\setminus C$ and $V_H \setminus C_H$ are independent sets of $G[M]$ and $H$, respectively. Moreover, $|M\setminus C| = |V_H \setminus C_H|$ since $F_v(G[M]) = F_v(H)$ and $F_{vc}(G[M]) = F_{vc}(H)$. According to Lemma \ref{vclemma}, the subgraph induced by $S \cup (M \setminus C)=V\setminus C$ in $G$ and the subgraph induced by $S \cup (V_H\setminus C_H)$ in $G'$ are isomorphic, so $S \cup (V_H\setminus C_H)$ is an independent set of $G'$. Hence, $(C\setminus M)\cup C_H$ is a VC of $G'$.

We claim $F_{cvc}(G') \leq F_{cvc}(G)$.
According to Lemma \ref{octlemma2}, $M\cap R$ is either $M$, a min-VC of $G[M]$, or $\{v\}$.
Suppose $M\cap R = M$. Clearly, $(R\setminus M) \cup V_H$ is a connected vertex cover (CVC) of $G'$.
Suppose $M\cap R$ is a min-VC of $G[M]$. Let $S = V \setminus (R\cup M)$. According to Lemma \ref{vclemma}, $G'[S \cup (V_H\setminus C_H)]$ and $G[S \cup (M \setminus R)] = G[V\setminus R]$ are isomorphic. Hence, $S \cup (V_H\setminus C_H)$ is an independent set of $G'$, and $(R\setminus M)\cup C_H$ is a VC of $G'$. Clearly, $G'[(R\setminus M)\cup C_H]$ is connected since $G[R]$ is connected. Therefore, $(R\setminus M)\cup C_H$ is a CVC of $G'$.  
Suppose $M \cap R = \{v\}$. Assume $u\in M\setminus R$ (the case $M=\{v\}$ has been discussed). Since $G$ is connected, $M' \subseteq N(u)$. Thus, $M'\subseteq R$. Since $F_{vc}(G[M])=F_{vc}(H)$, $F_{v}(G[M])=F_{v}(H)$, and $\{v\}$ is a VC of $G[M]$, there exists $w\in V_H$ that covers all edges of $H$. Therefore, $(R\setminus \{v\})\cup \{w\}$ is a VC of $G'$. In addition, according to Lemma \ref{vclemma}, $G[R]$, which is connected, and the subgraph induced by $(R\setminus \{v\})\cup \{w\}$ in $G'$ are isomorphic. Thus, $(R\setminus \{v\})\cup \{w\}$ is a CVC of $G'$.

%Next, suppose $G$ is not connected. The case is trivial when $G$ contains only trivial connected components. Assume $G$ includes two non-trivial connected components. Then, the CVCs of $G$ do not exist and $F_{cvc}(G)=|V|+1$. Thus, $F_{cvc}(G')\leq |V'|+1 = F_{cvc}(G)$ since $|V'|=|V|$. Assume $G$ includes exactly one non-trivial connected component $K$ and other singletons. Clearly, $R \subseteq V(K)$. Suppose $M\subseteq V(K)$. Then the proof goes the same way that $G$ is connected. Suppose $M$ includes a trivial connected component and no vertices of $K$. Then $H$ has no edges as $F_{vc}(G[M])=F_{vc}(H)$. Thus, $R$ is a CVC of $G'$. Suppose $M$ contains a trivial connected component and a vertex of $K$. Then $V(K) \subseteq M$, otherwise, $M$ is not a module of $G$. This means that $G[V\setminus M]$ are trivial connected components. Thus, $F_{cvc}(G)=F_{cvc}(G[M]) = F_{cvc}(H)=F_{cvc}(G')$. 

We claim $F_{fvs}(G') \leq F_{fvs}(G)$. Based on Lemma \ref{octlemma2}, $F\cap M$ is either $M$, $M \setminus \{v\}$, a min-VC of $G[M]$, or a min-FVS of $G[M]$. Let $S= V\setminus (F\cup M)$. Clearly, $(F\setminus M) \cup V_H$ is an FVS of $G'$ if $F\cap M = M$.
Suppose $F\cap M = M \setminus \{v\}$. Let $u\in V_H$. Based on Lemma \ref{vclemma}, $G[S\cup \{v\}]$ and $G'[S\cup \{u\}]$ are isomorphic. Hence, $G'[S\cup \{u\}]$ is a forest, and $(F\setminus M) \cup (V_H\setminus \{u\})$ is an FVS of $G'$.
Suppose $F\cap M$ is a min-VC of $G[M]$. Based on Lemma \ref{vclemma}, $G[S\cup (M\setminus F)]=G[V\setminus F]$ and $G'[S \cup (V_H\setminus C_H)]$ are isomorphic. So  $G'[S \cup (V_H\setminus C_H)]$ has no cycles, and $(F\setminus M)\cup C_H$ is an FVS of $G'$.
Suppose $F\cap M$ is a min-FVS of $G[M]$. Assume $G[M\setminus F]$ has no edges. Based on Lemma \ref{noedge}, $G[M]$ has no edges. Thus, $H$ has no edges, moreover, $G$ and $G'$ are isomorphic according to Lemma \ref{vclemma}. Assume $G[M\setminus F]$ has an edge. Then $M'\subseteq F$. Hence, $(F\setminus M)\cup F_H$ is an FVS of $G'$.

We claim $F_{oct}(G') \leq F_{oct}(G)$. Based on Lemma \ref{octlemma2}, $O\cap M$ is either $M$, a min-VC of $G[M]$, or a min-OCT of $G[M]$. Let $S=V\setminus (O\cup M)$. Clearly, $(O\setminus M) \cup V_H$ is an OCT of $G'$ if $O\cap M = M$.
Assume $O\cap M$ is a min-VC of $G[M]$. According to Lemma \ref{vclemma}, $G[S\cup (M\setminus O)] = G[V\setminus O]$ and $G'[S \cup (V_H \setminus C_H)]$ are isomorphic. Therefore, $G'[S \cup (V_H \setminus C_H)]$ has no odd cycles, and $(O\setminus M)\cup C_H$ is an OCT of $G'$.
Assume $O\cap M$ is a min-OCT of $G[M]$. Suppose $G[M\setminus O]$ has no edges. Based on Lemma \ref{noedge}, $G[M]$ has no edges. Thus, $H$ has no edges, moreover, $G$ and $G'$ are isomorphic according to Lemma \ref{vclemma}. Suppose  $G[M\setminus O]$ has an edge, then $M'\subseteq O$. Hence, $(O\setminus M)\cup O_H$ is an OCT of $G'$.
\end{proof}

A partition of a graph $I$ into paths is a set of vertex disjoint paths of $I$ whose union contains all vertices of $V(I)$. 
A connected subgraph of a path $L$ is called a sub-path of $L$. 
The operation of substituting a path $P$ for the sub-path $L'$ of $L$ is to first delete $L'$ from $L$, and then connect the two endpoints of $P$ with the two cut endpoints of $L - L'$, respectively.

\begin{lemma}
\label{partition into paths}
\textsc{Partitioning Into Paths($mw$)} and \textsc{Hamiltonian Cycle($mw$)} have PTCs.
\end{lemma}
\begin{proof}
For any $I \in \mathcal{G}$, assume function $c(I)$ denotes the number of the connected components of $I$, function $F_{pip} (I)$ denotes the smallest integer $k$ such that $I$ has a partition into $k$ paths, and function $F_{hc} (I)=0$ if $I$ has a Hamiltonian cycle, otherwise $F_{hc} (I)=1$.  
Suppose $F_{hc} (G[M])=F_{hc} (H)$, $F_{pip} (G[M])=F_{pip} (H)$, and $F_{v} (G[M])=F_{v} (H)$. Clearly, $F_{v} (G')\leq F_{v} (G)$.

Assume $\mathcal{L}$ is a partition of $G$ into $F_{pip} (G)$ paths. Assume $\mathcal{O} = \{L \in \mathcal{L} \mid  V(L) \subseteq  V\setminus M \}$, $\mathcal{Q} = \{L \in \mathcal{L} \mid 
V(L) \subseteq  M \}$, and $\mathcal{P} = \mathcal{L} \setminus (\mathcal{O}\cup \mathcal{Q})$. 
Suppose $s$ equals the sum of $c(L\cap G[M])$ for all $L \in \mathcal{P}$. Then $F_{pip} (G[M]) \leq s + |\mathcal{Q}| \leq F_{v} (G[M])$. Hence, $F_{pip} (H) \leq s + |\mathcal{Q}| \leq F_{v} (H)$, and there exists a partition of $H$ into paths $\mathcal{R}$ with $s + |\mathcal{Q}|$ elements. 
Now, construct a partition of $G'$ into paths $\mathcal{L'}$ as follows.  (1) Add all elements of $\mathcal{O}$ into $\mathcal{L'}$. 
(2) Do the following process for every $L\in \mathcal{P}$. Suppose $L\cap G[M]$ consists of connected components $L_1,\ldots,L_r$. Clearly, each $L_i$ is a sub-path of $L$ for $i\in [r]$. For every $L_i$, substitute a path of $\mathcal{R}$ for the sub-path $L_i$ in $L$, and then delete the path from $\mathcal{R}$. After substituting for all the $r$ sub-paths in $L\cap G[M]$, we obtain a new path $L'$ of $G'$ and add $L'$ to $\mathcal{L'}$. 
(3) Add all remaining paths of $\mathcal{R}$ into $\mathcal{L'}$. 
Clearly, $|\mathcal{O}|$ paths are added into $\mathcal{L'}$ in the first step and $|\mathcal{P}|$ paths are added into $\mathcal{L'}$ in the second step. Since $s$ paths are deleted from $\mathcal{R}$ in the second step, $|\mathcal{R}| - s = |\mathcal{Q}|$ paths are added into $\mathcal{L'}$ in the third step. Thus, $|\mathcal{L'}| = |\mathcal{L}|$ and $F_{pip} (G')\leq F_{pip} (G)$.

Suppose $C$ is a Hamiltonian cycle of $G$. Then $c(C\cap G[M])=c(C\cap G[V\setminus M])$. Since $F_{pip} (G[M])=F_{pip} (H)$ and $F_{v} (G[M])=F_{v} (H)$, there exists a partition of $H$ into paths with $c(C\cap G[M])$ elements. Applying the similar substitution operations on $C$ as that on $L\in \mathcal{P}$ in the proof for $F_{pip} (G')\leq F_{pip} (G)$, a Hamiltonian cycle of $G'$ can be obtained. Thus, a Hamiltonian cycle of $G$ implies a Hamiltonian cycle of $G'$, and $F_{hc} (G')\leq F_{hc} (G)$.
\end{proof}

\begin{lemma}
\label{longest induced path ptc}
\textsc{Longest Induced Path($mw$)} has a PTC.
\end{lemma}
\begin{proof}
Let function $F (I)$ represent the vertex number of the longest induced path for any $I\in \mathcal{G}$. Suppose w.l.o.g. $F (I) \geq 4$ (otherwise it is polynomial-time solvable). 
Suppose $F (G[M])=F (H)$.
Assume $L$ is a longest induced path of $G$.
Then, $|N(v)\cap V(L)| \leq 2$ for any $v\in V(L)$.
%Then, any vertex of $V(L)$ has at most two adjacent vertices of $V(L)$ in $G$. 
First, assume $V(L) \cap M =\emptyset$. Clearly, $L$ is also an induced path of $G'$, so $F (G)\leq F (G')$. 
Secondly, assume $V(L) \cap M = \{u\}$. Then, for any $v\in V_H$, since $v$ and $u$ have the same neighborhood in $M'$, the subgraph induced by $(V(L) \setminus \{u\})\cap \{v\}$ in $G'$ is an induced path with $|V(L)|$ vertices. Thus, $F (G)\leq F (G')$. 
Thirdly, assume $V(L) \cap M = \{u,v\}$. Since $|V(L)|\geq 4$, $|V(L)\cap M'| \geq 1$. If $V(L)\cap M'$ contains exactly one vertex, say $w$, then there exists $x \in V(L) \cap (V \setminus N[M])$ such that $wx \in E(L)$. So $wu, wv, wx \in E(L)$, a contradiction. If $V(L)\cap M'$ contains at least two vertices, say $w,x$, then the subgraph induced by $\{u, v, w, x\}$ contains a $C_4$, a contradiction. 
Fourthly, assume $3 \le |V(L) \cap M| < |V(L)|$. Then there exists $v\in V(L) \cap M'$ such that $|N(v)\cap V(L)|\geq 3$, a contradiction.
Finally, assume $V(L) \subseteq M$. Since $F (G[M])=F (H)$, $H$ has an induced path with $|V(L)|$ vertices, so $F (G)\leq F (G')$.
\end{proof}

%Let the independent triangle packing number of a graph be the number of triangles of a maximum independent triangle packing of the graph. Furthermore, we define the independent cycle packing number and the induced matching number in the analogous way.

\begin{lemma}
\label{induced matching}
\textsc{Induced Matching($mw$)}, \textsc{Independent Triangle Packing($mw$)}, and \textsc{Independent Cycle Packing($mw$)} have PTCs.
\end{lemma}
\begin{proof}
Suppose functions $F_{vc}(K)$, $F_{im}(K)$, $F_{itp}(K)$, and $F_{icp}(K)$ denote the element numbers of the min-VC, the max-IM, max-ITP, and max-ICP of any $K \in \mathcal{G}$, respectively. Suppose $F_{vc}(G[M])=F_{vc}(H)$, $F_{im}(G[M])=F_{im}(H)$, $F_{itp}(G[M])=F_{itp}(H)$, $F_{icp}(G[M])=F_{icp}(H)$, and $F_{v}(G[M])=F_{v}(H)$. Let graphs $I$, $T$, and $P$ represent a  max-IM, a max-ITP, and a max-ICP of $G$, respectively. 
Clearly, $F_{v}(G)\leq F_{v}(G')$. According to the proof of Lemma \ref{OCT}, $F_{vc}(G)\leq F_{vc}(G')$ since $F_{vc}(G[M])=F_{vc}(H)$ and $F_{v}(G[M])=F_{v}(H)$.
Recall that $M'\neq \emptyset$.

We claim $F_{im}(G)\leq F_{im}(G')$. Clearly, $F_{im}(G)=F_{im}(G[V\setminus M]) \leq F_{im}(G')$ if $V(I) \cap M = \emptyset$. Assume $V(I) \cap M$ consists of one vertex, say $v$. Then $V(I) \cap M' = \{u\}$ and $uv \in E(I)$. Hence, for any $w\in V_H$, $N(w)\cap V(I) = \{u\}$, and $G'[\{w\}\cup (V(I)\setminus\{v\})]$ is an induced matching of $G'$.
Assume $V(I)\cap M$ contains more than one vertex. Then, $V(I)\cap M' = \emptyset$. Thus, for any edge of $I$, both its endpoints are in either $M$ or $V\setminus N[M]$. Suppose $I_{G[M]} = I\cap G[M]$, and $I_H$ is a max-IM of $H$. Clearly, $I_H \cup (I - I_{G[M]})$ is an induced matching of $G'$, where $|E(I_H)|\geq |E(I_{G[M]})|$.

We claim $F_{itp}(G)\leq F_{itp}(G')$.
Clearly, $F_{itp}(G)=F_{itp}(G[V\setminus M])\leq F_{itp}(G')$ if $V(T) \cap M = \emptyset$. Assume $V(T)\cap M$ consists of one vertex, say $u$. Then $V(T)\cap M' = \{v,w\}$ and $G[\{u,v,w\}]$ is a triangle of $T$. Moreover, $N(x)\cap V(T) = \{v,w\}$ for any $x\in V_H$.  Thus, $G'[\{x\}\cup (V(T) \setminus \{u\})$ is an independent triangle packing (ITP) of $G'$. 
Assume $V(T)\cap M$ consists of two vertices, say $u, v$. If $u,v$ belong to different triangles of $T$, then at least four vertices in $V(T)\cap M'$ are adjacent to $u$ (or $v$), a contradiction. 
Assume $u,v$ are in the same triangle of $T$. Then $V(T)\cap M' = \{w\}$ and $G[\{u,v,w\}]$ is a triangle of $T$. 
Additionally, since $F_{im}(G[M])=F_{im}(H)$ and $G[M]$ contains $uv$, there exists an $xy\in E_H$. Thus, $G'[\{x,y\} \cup (V(T) \setminus \{u,v\})]$ is an ITP of $G'$.
Assume $V(T)\cap M$ includes at least three vertices. Then $V(T)\cap M' = \emptyset$. Thus, for every triangle of $T$, all its vertices are in either $M$ or $V\setminus N[M]$. Suppose $T_{G[M]} = T \cap G[M]$, and $T_H$ is a max-ITP of $H$. Then, $T_H\cup (T - T_{G[M]})$ is an ITP of $G'$, where the number of triangles of $T_H$ is at least that of $T_{G[M]}$.

We claim $F_{icp}(G)\leq F_{icp}(G')$. Clearly, $F_{icp}(G)=F_{icp}(G[V\setminus M])\leq F_{icp}(G')$ if $V(P) \cap M = \emptyset$. 
Suppose $V(P) \cap M$ consists of a vertex $u$. There exist $v, w\in M'$ such that $v,w \in V(C)$, where $C$ is a cycle of $P$. Moreover, $V(P)\cap M' = \{v,w\}$. Thus, $N(x)\cap V(P) = \{v,w\}$ for any vertex $x$ in $H$. Clearly, $G'[\{x\}$ $\cup$ $(V(P) \setminus \{u\})]$ is an independent cycle packing (ICP) of $G'$.  
Suppose $V(P)\cap M$ consists of two vertices $u,v$. If $u,v$ belong to different cycles of $P$, then at least four vertices in $V(P) \cap M'$ are adjacent to $u$ (or $v$), a contradiction.  
Assume $u,v$ are in the same cycle $C$ of $P$. Suppose $uv\in E(C)$. Then $M'\cap V(P)$ consists of one vertex, say $w$, otherwise, $|V(C)| < 3$ or $u$ (or $v$) has more than two neighbor vertices in $N[M]\cap V(P)$. Thus, $C$ is the triangle $G[\{u,v,w\}]$. In addition, there exists an $xy\in E_H$ since $F_{im}(G[M])=F_{im}(H)$ and $G[M]$ has an edge. Thus, $G'[\{x,y\} \cup (V(P) \setminus \{u,v\})]$ is an ICP of $G'$.
Suppose $uv\not\in E(C)$. Then $M'\cap V(P)$ consists of two vertices, say $w, z$, otherwise, $C$ is not a cycle or $u$ (or $v$) has more than two neighbor vertices in $N[M]\cap V(P)$. Moreover, $wz \not\in E$ since $w$ (or $z$) has two neighbours $u,v \in V(C)$. Thus, $C = G[\{u,v,w,z\}]$ is a cycle with four vertices.
Since $F_{v}(G[M])=F_{v}(H)$, $F_{vc}(G[M])=F_{vc}(H)$ and $\{u,v\}$ is an independent set of $G[M]$, there exists an independent set of $H$ with at least two vertices, say $x,y$. Thus, $G'[\{x,y\} \cup (V(P) \setminus \{u,v\})]$ is an ICP of $G'$.
Suppose $|V(P) \cap M|\geq 3$. Then $V(P)\cap M' =\emptyset$. Thus, for every cycle $C$ of $P$, all vertices of $C$ are in either $M$ or $V\setminus N[M]$. In addition, since $F_{icp}(G[M])=F_{icp}(H)$, the max-ICP of $G'$ is at least that of $G$. 
\end{proof}

Corollary \ref{corollary-in-all-ptcs} holds according to Lemma \ref{dominating set 222}, \ref{OCT}, \ref{partition into paths} and the following reasons. (1) An independent set, a nonblocker, and the vertex set of a maximum induced forest of $G$ are complements of a vertex cover, a dominating set,  and a feedback vertex set of $G$, respectively. (2) $mw$ does not change under graph complementation, and an independent set in $G$ is a clique in the complement graph of $G$. (3) The $mw$ of the output graph equals that of the input graph using the routine reduction from \textsc{Hamiltonian Path} to \textsc{Hamiltonian Cycle}.

\begin{corollary}\label{corollary-in-all-ptcs}
\textsc{Independent Set($mw$)}, \textsc{Clique($mw$)}, \textsc{Maximum Induced Forest($mw$)}, \textsc{Nonblocker($mw$)}, and \textsc{Hamiltonian Path($mw$)} have PTCs.
\end{corollary}

\section{Polynomial compression lower bounds for problems}

We provide polynomial compression (PC) lower bounds parameterized by $mw$ for the last six problems in Theorem \ref{theorem-ptc} using  the cross-composition technique \cite{DBLP:journals/siamdm/BodlaenderJK14}.

\begin{definition}[Polynomial equivalence relation \cite{DBLP:journals/siamdm/BodlaenderJK14}]
An equivalence relation $R$ on $\Sigma^*$ is called a \textit{polynomial equivalence relation} if the following two conditions hold: 
\begin{enumerate}
\item
there is an algorithm that given two strings $x,y\in \Sigma^*$ decides whether $x$ and $y$ belong to the same equivalence class in $(|x|+|y|)^{\Oh(1)}$ time;
\item
for any finite set $S \subseteq \Sigma^*$ the equivalence relation $R$ partitions the elements of $S$ into at most $(\max_{x\in S}|x|)^{\Oh(1)}$ classes.
\end{enumerate}
\end{definition}

\begin{definition}[And-cross-composition (or-cross-composition) \cite{DBLP:journals/siamdm/BodlaenderJK14}]
Let $L\subseteq \Sigma^*$ be a set and let $Q\subseteq \Sigma^* \times \mathbb{N}$ be a parameterized problem. We say that $L$ \textit{and-cross-composes} (\emph{or-cross-composes}) into $Q$ if there is a polynomial equivalence relation $R$ and an algorithm which, given $t$ strings $x_1,\ldots,x_t$ belonging to the same equivalence class of $R$, computes an instance $(y,k)\in \Sigma^* \times \mathbb{N}$ in time polynomial in $\sum_{i=1}^{t} |x_i|$ such that: 
%(1) $(y,k)\in Q$ if and only if all instances $x_i$ are yes for $L$ (at least one instance $x_i$ is yes for $L$); (2) $k$ is bounded by a polynomial in $\max_{i=1}^{t} |x_i|+\log t$.
\begin{enumerate}
\item 
$(y,k)\in Q$ if and only if all instances $x_i$ are yes for $L$ (at least one instance $x_i$ is yes for $L$);
\item 
$k$ is bounded by a polynomial in $\max_{i=1}^{t} |x_i|+\log t$.
\end{enumerate}
\end{definition}

\begin{theorem}[\cite{DBLP:journals/siamdm/BodlaenderJK14}]
Let $L$ be an NP-hard problem under Karp reductions. If $L$ and/or-cross-composes into a parameterized problem $Q$, then $Q$ does not admit a PC unless NP $\subseteq$ coNP/poly.
\end{theorem}

Polynomial compression (PC) lower bounds for the first 11 problems of Theorem \ref{theorem-ptc} are provided in \cite{Tablepaper} that include \textsc{Dominating Set}$(mw)$, \textsc{Feedback Vertex Set}$(mw)$, and \textsc{Hamiltonian Path}$(mw)$, so PC lower bounds for \textsc{Nonblocker}$(mw)$, \textsc{Maximum Induced Forest}$(mw)$, and \textsc{Partitioning Into Paths}$(mw)$ are obtained immediately. Next, we use the cross-composition \cite{DBLP:journals/siamdm/BodlaenderJK14} to prove \textsc{Longest Induced Path}$(mw)$, \textsc{Independent Triangle Packing}$(mw)$, and \textsc{Independent Cycle Packing}$(mw)$ have no PCs unless NP $\subseteq$ coNP/poly. 
For each and/or-cross-composition from a problem $L$ to a problem $Q$ and its related polynomial equivalence relation $R$ on $\Sigma^*$ in this section, there will be a bad equivalent class under $R$, which consists of invalid instances of $L$ that can be handled trivially, so we may assume $\Sigma^*$ only includes valid instances of $L$.

%We define the size of an independent triangle packing of a graph is the number of triangles in the independent triangle packing. \textsc{Independent triangle packing} is NP-complete \cite{DBLP:journals/mp/CameronH06}. Next, 

We define a new problem \textsc{Independent Triangle Packing Refinement} (ITPR)  as follows: the input is a graph $G$ and an independent triangle packing (ITP) of $G$ with $k$ triangles, decide whether $G$ has an ITP with $k+1$ triangles?

\begin{lemma}
\label{ITP-nph}
ITPR is NP-hard under Karp reductions.
\end{lemma}
%\begin{proofsketch}
%We provide a Karp reduction from \textsc{Independent Triangle Packing} to ITPR. Given an instance $(G,k)$, where $G=(V,E)$ and $V=\{v_1,\ldots,v_n\}$. Assume w.l.o.g. that $2 \leq k \leq \frac{n}{3}$. Construct $G'=(V',E')$ as follows. First, add $G$, vertices $x_1,\ldots,x_{n-k+1}$, and $n$ triangles $u_1$-$w_1$-$w'_1,\ldots ,u_n$-$w_n$-$w'_n$ into $G'$. Then, connect $u_i$ with all vertices of $V$ for each $i\in [n]$. Finally, connect $x_i$ with $w_i, w'_i$ for each $i\in [n-k]$ and connect $x_{n-k+1}$ with $w_{n-k+1},\ldots,w_n,$ $w'_{n-k+1},\ldots,w'_n$. Clearly, $T=\{u_1$-$w_1$-$w'_1,\ldots, u_n$-$w_n$-$w'_n\}$ is an ITP of $G'$. Thus, $(G',T)$ is an instance of ITPR with $n$ triangles. Then, we can prove that $(G,k)$ is a yes instance of \textsc{Independent Triangle Packing} if and only if $(G',T)$ is a yes instance of ITPR.
%\end{proofsketch}
\begin{proof}
Let $u$-$v$-$w$ denote the triangle induced by vertices $u,v,w$ of a graph. Provide a Karp reduction from \textsc{Independent Triangle Packing} to ITPR. Given an instance $(G,k)$, where $G=(V,E)$ and $V=\{v_1,\ldots,v_n\}$. Assume w.l.o.g. that $2 \leq k \leq \frac{n}{3}$ (otherwise, $(G,k)$ can be decided in polynomial time). 
Construct a graph $G'=(V',E')$ as follows. First, add $G$, vertices $x_1,\ldots,x_{n-k+1}$, 
and $n$ triangles $u_1$-$w_1$-$w'_1,\ldots ,u_n$-$w_n$-$w'_n$ into $G'$. Then, connect $u_i$ with all vertices of $V$ for each $i\in [n]$. Finally, connect $x_i$ with $w_i, w'_i$ for each $i\in [n-k]$ and connect $x_{n-k+1}$ with $w_{n-k+1},\ldots,w_n,$ $w'_{n-k+1},\ldots,w'_n$. Clearly, $T=\{u_1$-$w_1$-$w'_1,\ldots, u_n$-$w_n$-$w'_n\}$ is an ITP of $G'$. Thus, $(G',T)$ is an instance of ITPR with $n$ triangles.
	
Assume $(G,k)$ is a yes instance of \textsc{Independent Triangle Packing}. Then, the $k$ independent triangles in $G$ together with the $n-k+1$ independent triangles $x_1$-$w_1$-$w'_1,\ldots,x_{n-k+1}$-$w_{n-k+1}$-$w'_{n-k+1}$ compose an ITP of $G'$ with size $n+1$. 
For the other direction, assume $(G',T)$ is a yes instance. Then $G$ has an ITP $T'$ with $n+1$ triangles. Let $U =\{u_1,\ldots,u_n\}$. We claim $V(T') \cap U = \emptyset$. Assume $V(T')\cap U = \{u_i\}$ for some $i\in [n]$. Then $N(u_i) \cap V(T')$ consists of two vertices that are either $w_i,w'_i$ or $u,v$, where $uv\in E$. 
Assume $N(u_i) \cap V(T')=\{u,v\}$. Then $V(T') \cap N(\{u_i,u,v\})= \emptyset$. Therefore, $V(T')$ is a subset of $L=V'\setminus N(\{u_i,u,v\})$, and the max-ITP of $G'[L]$ equals that of $G'$. Clearly, the size of the max-ITP of $G'[L]$ is $n-k+1 < n$ if $i\in [n-k]$, and is $n-k+2 \leq n$ if $n-k+1\leq i \leq n$, a contradiction. 
Assume $N(u_i) \cap V(T')= \{w_i,w'_i\}$. Then $V(T') \cap N(\{u_i,w_i,w'_i\}) = \emptyset$. Therefore, $V(T')$ is a subset of $L=V'\setminus N(\{u_i,w_i,w'_i\})$, and the max-ITP of $G'[L]$ equals that of $G'$. Clearly, the size of the max-ITP of $G'[L]$ is $n$, a contradiction.
Now, we have $V(T') \subseteq V'\setminus U$. Clearly, the size of the max-ITP of $G'[V'\setminus (U\cup V)]$ is $n-k+1$. Thus, there is an ITP of $G$ with size at least $(n+1)-(n-k+1)=k$.
\end{proof}

\begin{lemma}
\label{ITP-no-polykernel}
\textsc{Independent Triangle Packing($mw$)} has no PCs unless NP $\subseteq$ coNP/poly.
\end{lemma}
%\begin{proofsketch}
%We can provide an or-cross-composition from ITPR to it, where the output instance is the disjoint union of the input instances.  
%\end{proofsketch}
\begin{proof}
Let function $F_{itp} (I)$ be the element number of the largest independent triangle packing of any $I\in \mathcal{G}$.
We provide an or-cross-composition from ITPR to \textsc{Independent Triangle Packing($mw$)}. Assume any two instances $(G_1,T_1)$ and $(G_2,T_2)$ of ITPR are equivalent under $R$ if and only if $|V(G_1)|=|V(G_2)|$, $|V(T_1)|=|V(T_2)|$, and $mw(G_1)=mw(G_2)$. Clearly, $R$ is a polynomial equivalence relation. 
Consider the or-cross-composition. Given $t$ instances $(G_1,T_1),$ $\ldots,$ $(G_t,T_t)$ of ITPR in an equivalence class of $R$, where $|V(G_i)|=n$, $V(T_i) = 3k$, and $mw(G_i) = l$ for all $i\in [t]$. Produce $(G, kt+1)$ in $\Oh(tn^2)$ time, where $G = \bigcup_{i=1}^{t}G_i$. Clearly, $mw(G) = l \leq n$ and $f_{itp} (G)=\Sigma_{i=1}^{t} f_{itp} (G_i)$. Thus, $f_{itp} (G_i)\geq k+1 $ for at least one $G_i$ if and only if $f_{itp} (G) \geq kt + 1$. 
\end{proof}

We define a new problem named \textsc{Independent Cycle Packing Refinement} (ICPR) as follows: the input is a graph $G$ and an independent cycle packing (ICP) of $G$ with $k$ cycles, decide whether $G$ has an ICP with $k+1$ cycles?

\begin{lemma}
\textsc{Independent Cycle Packing($mw$)} has no PCs unless NP $\subseteq$ coNP/poly.
\end{lemma}
\begin{proof}
That ICPR is NP-hard under Karp reductions can be proved in the same way as that of Lemma \ref{ITP-nph}. Then, that \textsc{Independent Cycle Packing($mw$)} has no PCs unless NP $\subseteq$ coNP/poly can be proved in the same way as that of Lemma \ref{ITP-no-polykernel}.
\end{proof}

\begin{lemma}
\textsc{Longest Induced Path($mw$)} has no PCs unless NP $\subseteq$ coNP/poly.
\end{lemma}
\begin{proof}
Let function $F_{lip} (I)$ be the vertex number of the longest induced path of any $I\in \mathcal{G}$. We provide an or-cross-composition from \textsc{Longest Induced Path} (LIP) to \textsc{Longest Induced Path($mw$)}. Assume any two instances $(G_1,k_1)$ and $(G_2,k_2)$ of LIP are equivalent under $R$ if and only if $|V(G_1)|=|V(G_2)|$ and $|k_1|=|k_2|$. Clearly, $R$ is a polynomial equivalence relation. 
Consider the or-cross-composition. Given $t$ instances $(G_1,k),\ldots,(G_t,k)$ of LIP in an equivalence class of $R$, where $|V(G_i)|=n$ for all $i\in [t]$. Produce $(G',k)$ in $\Oh(tn^2)$ time, where $G' = \bigcup_{i=1}^{t}G_i$. Clearly, $mw(G') = \max_{1\leq i\leq t} mw(G_i) \leq n$ and $F_{lip} (G')=\max_{1\leq i\leq t} F_{lip} (G_i)$. Thus, at least one $G_i$ with $F_{lip} (G_i)\geq k$ if and only if $F_{lip} (G')\geq k$.  
\end{proof}

%\section{Conclusions}
%In this paper we try to explore more examples for polynomial Turing compression. We find that when we solve a graph problem by using the modular decomposition tree of the graph, we need to investigate the quotient graphs of the vertices of the tree repeatedly. This mode is similar with the way that Turing compression solves problems. Based on this idea, polynomial Turing compressions are demonstrated for several graph problems parameterized by modular-width. Moreover, we construct a recipe to facilitate the process of obtaining these results. Additionally, we find conditional tight kernels for a few graph problems parameterized by modular-width, and new parameterized algorithms for these problems are obtained.

%\noindent \textbf{Conclusions.} 
\section{Conclusions}
We conclude this paper by proposing some open questions.
Does \textsc{$k$-Path} have a PTC parameterized by $mw$? In addition, Fomin et al. \cite{Fomin_2017} gives a meta-theorem that proves a family of problems is FPT parameterized by $mw$. Can we also give a meta-theorem to prove the problems in that family have PTCs? 

\vspace{2mm}

\section*{Acknowledgements}
%\noindent \textbf{Acknowledgements.} 
I thank Manuel Lafond for his careful reading and constructive comments to improve this manuscript, as well as his valuable help in many other aspects. I thank the anonymous referees for their valuable comments on the improvement of this manuscript. This work is supported by CSC 201708430114 fellowship.


\bibliographystyle{splncs04}
\bibliography{mybibliography}

\end{document}